\documentclass[prc,aps,preprint,showpacs]{revtex4}
\usepackage{graphicx}

\begin{document}
\title{Possibile Presence and Properties of Multi Chiral Pair-Bands in Odd-Odd
Nuclei with the Same Intrinsic Configuration} 

\author{ Ikuko Hamamoto$^{1,2}$ }

\affiliation{
$^{1}$ {\it Riken Nishina Center, Wako, Saitama 351-0198, Japan } \\ 
$^{2}$ {\it Division of Mathematical Physics, Lund Institute of Technology 
at the University of Lund, Lund, Sweden} }   


  

\begin{abstract}
Applying a relatively simple particle-rotor model to odd-odd nuclei, 
possible presence of multi chiral pair-bands
is looked for, where chiral pair-bands  
are defined not only by near-degeneracy 
of the levels of two bands but also by almost the same expectation values of 
squared components of three  
angular-momenta that define chirality.
In the angular-momentum region where two pairs of chiral pair-bands are
obtained the possible interband 
M1/E2 decay from the second-lowest chiral pair-bands 
to the lowest
chiral pair-bands is studied, with the intention of finding 
how to experimentally
identify the multi chiral pair-bands.  It is found that up till almost band-head
the intraband M1/E2 decay within
the second chiral pair-bands is preferred rather than the interband 
M1/E2 decay to 
the lowest chiral pair-bands, though the decay possibility depends on the ratio
of actual decay energies.  
It is also found that chiral pair-bands  
in our model and definition are hardly
obtained for $\gamma$ values outside the range 
$25^{\circ} < \gamma < 35^{\circ}$,
although either 
a near-degeneracy or a constant energy-difference of several hundreds
keV between the two levels for a given angular-momentum $I$ 
in ''a pair bands'' is sometimes obtained in some
limited region of $I$.  
In the present model calculations the energy difference 
between chiral pair-bands is always one or two orders of magnitude smaller
than a few hundreds keV, and no chiral pair-bands are obtained, 
which have an almost constant energy difference of the order of 
a few hundreds keV 
in a reasonable range of $I$.   

\end{abstract}


\maketitle

\newpage

\section{INTRODUCTION} 
The total Hamiltonian for the nuclear system is taken to be invariant under the
exchange of the right- and left-handed geometry.  Chirality in triaxial nuclei
is characterized by the presence of three angular-momentum vectors which are
noncoplaner and thereby make it possible to define chirality.  
Since possible triaxial even-even nuclei are generally expected to collectively 
rotate mainly about the 
intermediate axis (taken as the 2-axis in the present article) 
as is expected from irroational-flow-like moments of
inertia, other two angular-momenta to define chirality 
must come from particle configurations.   In odd-odd nuclei the simplest
example is the angular momenta of odd neutron and odd proton, 
which prefer to pointing out the directions of the shortest and 
longest axes.  

The occurrence of the chirality in a nuclear structure was considered
theoretically in \cite{FM97}, and since then experimental spectra 
exhibiting two $\Delta I$ = 1 rotational bands, which presumably 
have the same parity and 
an almost constant energy difference, have been
reported in the region of the mass number $A \approx$ 130 and 
110 region.  The usual interpretation is that in the $A \approx 130$ region a
proton-particle and a neutron-hole in the $h_{11/2}$-shell play a role in
producing the two
angular-momenta to define chirality, while in the $A \approx 110$ 
region a
proton-hole in the $g_{9/2}$-shell and a neutron-particle in the
$h_{11/2}$-shell play the role.  
The energy difference between the observed levels with the same $I$ 
belonging to those observed pair-bands is typically 
several hundreds keV in the $A \approx 130$ region where more data are reported, 
though it is not clear whether or not  
the observed difference is really 
close to a constant in the relevant angular-momentum region.    
It is not easy to find the origin which gives rise to 
such an amount of constant 
energy difference when the chiral pair-bands are realized.    

It has 
been theoretically known for years \cite{KS02,DRS09} that in some limited range
of $I$ multiple pair-bands, of which the two levels with a given $I$ 
are degenerate with very good accuracy, 
are obtained for $\gamma \sim 30^{\circ}$ and 
a given chiral-candidate configuration with one 
high-$j$ proton-particle and one high-$j$ neutron-hole.  In the 
present article we use 
the conventional way of
defining the triaxial parameter \cite{BM75}, 
$0^{\circ} \leq \gamma \leq 60^{\circ}$, 
which corresponds 
to the region $0^{\circ} \geq \gamma \geq -60^{\circ}$ in the Lund
convention \cite{GA76} employed conventionally in high-spin physics.  
A high-j orbit in a given major shell has unique properties such as a 
unique parity and a large angular-momentum compared with other
energetically close-lying one-particle 
orbits so that the high-j one-particle wave-functions 
remain relatively pure under both deformation and rotation and, furthermore, 
the states containing high-j particles appear close to the yrast line in high
angular-momenta.   
Therefore, it may be possible to observe higher-lying chiral
pair-bands consisting of the same high-$j_p$ quasiproton and high-$j_n$
quasineutron configulation as that of the lowest chiral pair-bands.  
Indeed, the present study was
prompted by the recent 
experimental finding of four (or five) very similar $\Delta
I$ = 1 bands with possibly the same parity in the odd-odd nucleus
$^{104}_{45}Rh_{59}$ \cite{JT13}, 
which may well be interpreted to come from the same
chiral-candidate configuration.
  
Pushing further the notion of chiral pair-bands, 
based on adiabatic and configuration-fixed constrained triaxial
relativistic mean field approaches the presence of multiple chiral pair-bands 
for different deformation 
($\beta$, $\gamma = 22^{\circ} \sim 31^{\circ} $) and 
different intrinsic configurations in a given nucleus 
$^{106}Rh$  was theoretically suggested \cite{MP06}, 
while in a recent publication \cite{AGA13} 
observed data on $^{133}_{58}Ce_{75}$  were interpreted in terms of two 
chiral doublet bands with positive and negative parity and different
deformations ($\beta$ = 0.20 $\sim$ 0.23, 
$\gamma$ = 11$^{\circ} \sim 15^{\circ}$).  
One may wonder whether it is possible to obtain chiral bands for a weak  
triaxial-deformation such as $\gamma = 15^{\circ}$, 
as the most favourable triaxial
deformation for realizing chirality is known to be  
$\gamma \approx 30^{\circ}$.

In the present work a relatively simple particle-rotor model 
of odd-odd nuclei consisting of a triaxial collective rotor together with 
one high-$j_p$ quasiproton and one high-$j_n$ quasineutron is used to study  
the possible presence and properties 
of two pairs of chiral bands as well as 
the relative structure of three angular-momenta, which define chirality.  
We would identify chiral pair-bands, only when not only the
near degeneracy of two $\Delta I$ = 1 bands but also the very similarity of the
expectation values of squared components of the three angular momenta in
some finite 
region of $I \gg 1$.  When the latter condition is fulfilled, the energies  of
the two bands as well as corresponding various intraband transitions 
are expected to be almost identical.   
In the case that the presence of two pairs of chiral
bands is obtained, the decay properties of the higher chiral pair-bands 
to the lower
chiral pair-bands are examined.    
First we study the case of a proton-particle and a neutron-hole in a given $j(=
h_{11/2})$-shell coupled to the collectively rotating core, since in this case a
quantum-number was present \cite{KSH04} 
for $\gamma = 30^{\circ}$ so that 
the understanding of numerical results is
transparent.
Then, we proceed to the case of a proton-hole in the
$g_{9/2}$-shell and a neutron-particle in the $h_{11/2}$-shell which may be
applicable to the possible pair-bands in the $A \approx$ 110 region.

In Sec.II main points of our model are briefly summarized, while numerical
results and discussions are presented in Sec.III.  Conclusions and discussions 
are given in Sec.IV.

\section{MODEL}
Our particle-rotor model Hamiltonian \cite{BM75} of odd-odd nuclei 
is written as 
\begin{equation}
H = H_{rot} + H_{intr}^{(p)} + H_{intr}^{(n)}
\label{eq:tH}
\end{equation}
where the first term on r.h.s. expresses the rotor Hamiltonian of 
the even-even core 
\begin{eqnarray}
H_{rot} & = & \sum_{k=1}^{3} \, \frac{\hbar^2}{2 \Im_k} \, R_k^2 \\
& = & \sum_{k=1}^{3} \, \frac{\hbar^2}{2 \Im_k} \, \left( I_k - (j_k^{(p)} + 
j_k^{(n)}) \right)^2
\end{eqnarray}
with the irrotationa-flow-like moments of inertia 
\begin{equation}
\Im_k = \frac{4}{3} \Im_0 \, \sin^2 (\gamma + k \frac{2 \pi}{3})
\label{eq:mt}
\end{equation}
The total angular-momentum $I$ is a good quantum-number, while the core 
angular-momentum $R$ is not.
In the present work we take the case in which 
both odd quasiproton and odd quasineutron are in respective single-$j$-shells. 
The intrinsic Hamiltonian in the case of a single-j-shell is conveniently
written as \cite{IH76,HM83}
\begin{equation} 
H_{intr} = \sum_{\nu} (\varepsilon_{\nu} - \lambda ) \, 
a_{\nu}^{\dagger} a_{\nu} 
+ \frac{\Delta}{2} \, \sum_{\mu, \nu} \, \delta(\bar{\mu}, \nu) 
\, (a_{\mu}^{\dagger}
a_{\nu}^{\dagger} + a_{\nu} a_{\mu})
\label{eq:spH}
\end{equation} 
where $\varepsilon_{\nu}$ expresses one-particle energies for a single-particle
with angular-momentum $j$ moving in a general triaxially-deformed quadrupole
potential 
\begin{equation}
V = \frac{\kappa}{j(j+1)} \left( \{ 3j_3^2 -j(j+1) \} \, \cos \gamma + \sqrt{3} \,
(j_1^2 - j_2^2) \, \sin \gamma  \right) . 
\label{eq:spV}
\end{equation}
In eq. (\ref{eq:spV}) 
$\kappa$ is used as a convenient energy unit for a single-$j$-shell, and the 
value of $\kappa$  is proportional to the quadrupole deformation parameter
$\beta$ \cite{BM75}.  An appropriate value of $\kappa$ may be something
between 2 and 2.5 MeV, depending on nuclei \cite{IH76}.
The quantities, $\varepsilon_{\nu}$, $\lambda$, $\Delta$ and $j$, 
appearing in $H_{intr}^{(p)}$ and $H_{intr}^{(n)}$ of (\ref{eq:tH}) depend on
protons or neutrons, respectively, 
while we take the values of $\kappa$
and $\gamma$ common to protons and neutrons.

Our particle-rotor Hamiltonian (\ref{eq:tH}) is numerically diagonalized in the
space consisting of the rotor coupled with one-quasiproton and one-quasineutron
which are obtained from the BCS approximation \cite{BM75}.  
The total number of basis states
for spin $I$ is given by $(1/2) \, (I+1/2) \,(2j_p+1) \,(2j_n+1)$ .

For M1 transitions the operator 
\begin{equation}
(M1)_{\mu} = \sqrt{\frac{3}{4 \pi}} \, \frac{e \hbar}{2mc} \, 
\{ (g_{\ell} - g_R) \, \ell_{\mu} + (g_s^{eff} - g_R) \, s_{\mu} \}
\end{equation}
is used.

When chiral geometry is realized, observed two states with $I$
in the chiral pair-bands may be written as \cite{KS01} 
\begin{equation}
\mid I+ \rangle =  \frac{1}{\sqrt{2}} (\mid IL \rangle + \mid IR \rangle ) 
\qquad \mbox{and} \qquad 
\mid I- \rangle = \frac{i}{\sqrt{2}} (\mid IL \rangle - \mid IR \rangle )
\end{equation}
where left- and right-handed geometry states are written as $\mid IL \rangle$
and $\mid IR \rangle$, respectively.  For states with $I \gg 1$ it is expected
that 
\begin{equation}
\langle IL \mid H \mid IR \rangle \approx 0, \qquad 
\langle IL \mid M1 \mid IR \rangle \approx 0, \qquad  
\mbox{and}  \qquad 
\langle IL \mid E2 \mid IR \rangle \approx 0 
\label{eq:ILIR}
\end{equation}
As $I$ increases the matrix-elements in (\ref{eq:ILIR}) rapidly decrease and 
may approach zero for the $I$-value at which chiral pair-bands 
are realized. 
If so, two levels with a given $I$ 
in the chiral pair-bands are almost degenerate, while 
mutually corresponding intraband M1 and E2 transitions in the chiral pair-bands
are almost identical.

Those energies and electromagnetic transitions can be, in principle, measured
experimentally, though the measurement of the absolute magnitude of the latter 
with good accuracy is 
at present not so easy.  
On the other hand, the occurresnce of chiral pair-bands can be
theoretically explored by examining the similarity of the 
three angular-momenta of the first band to
that of the second band, which define chirality.  
Checking the similarity is more fundamental and strict than
just exploring the degeneracy of levels for a given $I$.

The expectation values of squared components of three angular momenta that
define chirality were previously calculated in Ref.\cite{SQZ07}.  We define  
\begin{eqnarray}
R_i(I) & \equiv & \sqrt{\langle  I \mid R_i^2 \mid I \rangle} = 
\sqrt{\langle I \mid (I_i - j_{pi} - j_{ni})^2 \mid I \rangle} \nonumber \\
j_{pi}(I) & \equiv & \sqrt{\langle I \mid j_{pi}^2 \mid I \rangle} \nonumber \\
j_{ni}(I) & \equiv & \sqrt{\langle I \mid j_{ni}^2 \mid I \rangle}
\label{eq:rsq}
\end{eqnarray}
where $i$=1, 2 and 3. 
However, the similarity (or difference) of those expectation values in the
lowest two levels with a given $I$ was not really used to pin down 
the identification of chiral pair-bands.  
We study the quantities in (\ref{eq:rsq}) because 
when chiral pair-bands are
realized the quantities $R_i$, $j_{pi}$ and $j_{ni}$ must be very similar 
for the
pair-bands for a given $I$.  On the other hand, if those quantities are not very
similar, say different by more than 10 percent, in the two bands, 
they are not regarded as chiral pair-bands, even
if the two bands are nearly degenerate.

\section{NUMERICAL RESULTS}
We show numerical results, in which the parameters, $\lambda_p$, $\lambda_n$,
$\Delta_p$, $\Delta_n$, $\gamma$ and $\Im_0$ are taken to be independent of $I$. 
This is partly because only in a limited region of $I$ 
so-called ''chiral pair-bands'' are expected to occur 
and partly because we want to keep our
model as simple as possible.  In reality, those parameters may depend on $I$
even within the limited range of $I$,  
and
in a quantitative comparison with experimental data one may have to perform 
more elaborate calculations.  However, we think that 
if chiral pair-bands will ever appear, a simple schematic
model such as the present one should already indicate the basic elements.   

When we obtain the angular-momentum region, in which two pairs of chiral
bands are identified, we have calculated the interband 
M1 and E2 transitions between the
first pair-bands and the second pair-bands.  Then, it is found that the E2
transitions may not win against the M1 transitions for possible transition
energies.  Thus, only the calculated $B(M1)$ values 
are shown and discussed in the following.

In numerical calculations presented in this article we use $\Delta_p$ =
$\Delta_n$ = 0.1$ \kappa$, ($\Im_0 / \hbar^2$)$ \kappa$ = 70, $g_R$ =
0.4 and $g_s^{eff}$ = 0.7$g_s^{free}$.
In the following the numerical results for $\gamma$ = 30$^{\circ}$, in
which chiral pair-bands are most favorably produced, are mostly presented,
except at the end of subsection B where an example of a constant energy
difference between two bands is discussed.

\subsection{Both protons and neutrons in the $h_{11/2}$-shell}
In this subsection we consider the case that both protons and neutrons are in
the $h_{11/2}$-shell. In particular, one quasiproton represents almost one
particle in the $h_{11/2}$-shell while one quasineutron expresses 
almost one hole
in the $h_{11/2}$-shell.  This choice of Fermi levels together with $\gamma =
30^{\circ}$ is theoretically known to be most favorable for producing a pair of
chiral bands.  Moreover, if the condition, $\lambda_p = - \lambda_n$, is
fulfilled the quantum-number $A$ defined in Ref. \cite{KSH04} is a good quantum
number of the system.

First, in the case of the parameters that the quantum number $A$ in 
Ref. \cite{KSH04} is valid, 
we give a schematic sketch of possible $M1(I \rightarrow I-1)$
transitions between the two pairs of idealistic chiral pair-bands, ($f1$, $u1$) 
and ($f2$, $u2$).  
In Fig. 1 we show the allowed $M1(I \rightarrow I-1)$ transitions from
the second-lowest pair of chiral bands ($f2,u2$) to the lowest 
pair ($f1,u1$), which are
denoted by dotted-line arrows, as well as the allowed $M1(I \rightarrow
I-1)$ transitions within respective pairs that are 
expressed by solid-line arrows.
The quantum-number $A$ of each level is expressed by $\pm$ sign.  
In this example respective bands ($f1$, $u1$, $f2$ and $u2$) 
are defined so that $E2(I \rightarrow I-2)$ transitions are 
always allowed within a given band, 
though there may be  
some band-crossings within a given pair-bands (for example along the yrast line)
experimentally
observed.  It is noted that because the M1 operator contains a non-negligible
isoscalar component weak M1 transitions are possible between levels with the same sign of
the quantum-number $A$, but such weak M1 transitions are not drawn in Fig. 1.  
There are two kinds of M1-decay scheme between the first and the second 
chiral pair-bands.  
In Fig. 1a states belonging to the $f2$ band 
$M1(I \rightarrow I-1)$ decay always to
those in the $f1$ band, while states in the $u2$ band 
$M1(I \rightarrow I-1)$ decay to those in the $u1$ band.  
In contrast, in the case of Fig. 1b the states in the $f2$ ($u2$) band 
$M1(I \rightarrow I-1)$ decay to those in the $f1$ and $u1$ bands 
alternatively in $I$.
Whether or not M1 transitions denoted by dotted lines in Fig. 1 can be observed
depends on both the B(M1)-values and transition energies relative to those shown
by solid arrows.  
Examples of numerical values are shown later.  

We note that in Fig. 1 the feature of chiral pair-bands  
is represented by the
energy degeneracy of the two bands with the quantum-number 
$A = \pm$ for a given $I$ level 
as well as the equality of the $B(M1)$- and $B(E2)$-values of 
mutually corresponding intraband transitions.  
In contrast, the presence of the quantum-number $A$ and the resulting selection
rule in electromagnetic transitions come from the symmetry properties of the
Hamiltonian with the present parameters and not from the realization of
chiral pair-bands. 

In Fig. 2a calculated energies of the lowest four bands are shown for $\gamma =
30^{\circ}$ and the proton Fermi level $\lambda_p$ (neutron Fermi
level $\lambda_n$) placed 
on the lowest (highest) single-particle energies of the
$h_{11/2}$-shell. These parameters, $\gamma$, $\lambda_p$ and $\lambda_n$ 
together with a moderate amount of pair-correlation are known 
to be most favorable for producing chiral pair-bands. 
As is well known, chiral pair-bands may appear at best only in the region of
intermediate values of $I$. 
In the following presentation of our numerical results the name, 
$b1$, $b2$, ... 
are given for the sequence of the levels 
counting simply from the energetically lowest to the higher  
levels for a given $I$, and not a sequence of levels which are connected by
strong $E2(I \rightarrow I-2)$ transitions such as shown in Fig. 1.  
This naming is chosen because
in the practical 
presence of non-vanishing interactions between two bands it is not always
trivial to define a band, of which levels are connected by large  
$B(E2;I \rightarrow I-2)$ values, in contrast to the present case 
in which the quantum number $A$ 
can be used to define a sequence of levels in a band.  

In Fig. 2b calculated values of $R_1$ and $R_2$ for the lowest-lying two bands 
are shown,
while in Fig. 2c calculated values of $j_{pi}$ and $j_{ni}$ are plotted. 
Due to the construction of the model and the parameters used here, 
the values of $R_1$ are equal to those of $R_3$.  
From Fig. 2c it
is seen that the quantities $j_{pi}$ and $j_{ni}$ are not very efficient  
for showing the character of chiral pair-bands, because both the variation
in $I$ and the dependence on bands are relatively minor. 
From Figs. 2a, 2b and 2c one may safely state that the lowest two bands, 
$b1$ and $b2$,
form chiral pair-bands in the region of $16 \leq I \leq 24$.
Indeed it is amazing to see that in the limited region of $I$ the lowest two
bands, $b1$ and $b2$, form an almost perfect chiral pair-bands.   
Examining calcualted
$R_i$-values of the third- and fourth-lowest bands shown in Fig. 2d, 
together with energies
exhibited in Fig. 2a, we may say that the bands, $b3$ and $b4$, 
may form chiral pair-bands in the region of $18 \leq I \leq 24$.

In Fig. 3a examples of 
calculated $B(M1;I \rightarrow I-1)$ values both between the bands 
$b1$ and $b2$ and between the bands $b3$ and $b4$ are shown 
in the angular-momentum
region, where the character of respective pair-bands is confirmed above to be
chiral.
The parameter set used in this numerical example is the one, in which the
quantum number $A$ defined in Ref. \cite{KSH04} should work exactly. 
Therefore, the zigzag behavior of $B(M1;I \rightarrow I-1)$ values 
sketched qualitatively  
in Figs. 1a or 1b appears precisely, except at $I$=20 and
21 where a band-crossing occurs within both pairs of chiral bands, 
[$b1$ and $b2$] 
and [$b3$ and $b4$], though the band-crossings are somewhat 
difficult to be recognized in the scale of Fig. 2a.  
The occurrence of the band-crossing may be understood in terms of a very small
difference of 
the effective moments of inertia between the lowest (third lowest) and the
second lowest (fourth lowest) bands, which originates
from a very small matrix-element $<IL|H|IR> \neq 0$.  

In Fig. 3b we plot examples of 
calculated $B(M1;I \rightarrow I-1)$ values of the M1
transitions from the second chiral pair-bands to the first ones.  
The M1 transitions are seen to be again controlled by the quantum number $A$. 
However, the point here is that the B(M1)-values in Fig. 3b are at least about a
factor 50 smaller than those of competing M1 transitions shown 
in Fig. 3a, due to the different  
structure of three vectors, $\vec{R}$, $\vec{j_p}$, $\vec{j_n}$, 
in the first and
second chiral pair-bands.  
That means, when the second chiral pair-bands are
formed in addition to the first ones, the decays of the former will occur within
the pair-bands to the levels of the band-head and not to the first 
chiral pair-bands, if for a given initial state 
the ratio of the decay energy of the M1 transition shown in Fig. 3b  
to that shown in Fig.3a 
is not larger than by a factor of about $(50)^{1/3} \approx 3.7$.  
Such a large ratio of the decay energies is hardly obtained from numerical
examples of the present model such as those shown in Fig. 2a. 

Though we have tried various numerical calculations varying the parameters,
$\lambda_p$, $\lambda_n$, $\Delta_p$, $\Delta_n$, $\gamma$ and $\Im_0$, 
it is found that outside the region of 
$25^{\circ} < \gamma < 35^{\circ}$ chiral pair-bands defined in the present
work are hardly obtained.  It is also found that we could not obtain any chiral
pair-bands, of which the energy difference is nearly a constant of the order of 
a few hundreds keV in a certain range of $I$.

\subsection{Protons in the $g_{9/2}$-shell and neutrons in the $h_{11/2}$-shell}
First we consider the case that for $\gamma = 30^{\circ}$ one quasiproton
expresses almost one hole in the $g_{9/2}$-shell while one quasineutron
represents almost one particle in the $h_{11/2}$-shell.
This choice of parameters is again most favorable for producing a pair of chiral
bands. 

In Fig.4a calculated energies of the lowest four bands are shown for $\lambda_p$
placed on the highest single-particle energy of the $g_{9/2}$-shell and for
$\lambda_n$ on the lowest single-particle energy of the $h_{11/2}$-shell.  The
behavior of the degeneracy of calculated energies of the lowest pair bands, 
$b1$ and $b2$, and the second lowest pair bands, $b3$ and $b4$, 
looks very similar to that in Fig. 2a.  

In Fig. 4b calculated values of $R_i$ of the lowest-lying two bands are shown.
From Fig. 4a and 4b one may state that the lowest-lying two bands, 
$b1$ and $b2$,
form chiral pair-bands in the region of $15 \leq I \leq 26$.  Examining
calculated $R_i$ values of the third- and fourth-lowest bands, 
$b3$ and $b4$, shown
in Fig. 4c, together with energies exhibited in Fig. 4a, we may say that the
bands, $b3$ and $b4$, 
form chiral pair-bands in the region of $17 \leq I \leq
22$.  It is noted that the appearance of chiral pair-bands, both the lowest pair
and the second lowest pair, in a certain range of $I$ is very similar to the
case that both protons and neutrons are in the $h_{11/2}$-shell discussed in the
previous subsection.  We also note that in the present model the character of
chiral pair-bands becomes dubious before the calculated energy difference
between two bands approaches 100 keV.  

In Fig.5a examples of calculated $B(M1;I \rightarrow I-1)$ values within the
bands $b1$ and $b2$ as well as 
within the bands $b3$ and $b4$ are plotted in the
angular-momentum region, where the chiral character of respective
pair-bands is identified.
It is seen that a strong zigzag pattern of the $B(M1)$-values, which is
reminiscent of Fig. 3a, appears especially for lower $I$-values.  The zigzag
pattern diminishes as $I$ increases.

In Fig. 5b calculated $B(M1;I \rightarrow I-1)$ values of 
the M1 transitions from
the second chiral pair-bands to the first ones are plotted.  It is again seen
that the B(M1)-values in Fig. 5b are at least about a factor 50 smaller that
those in Fig. 5a.  That means, when the second chiral pair-bands are formed in
addition to the first ones, the M1 decays of the levels of the former to those
of the latter will hardly occur except possibly around the bandhead, though the
decay possibility depends on actual transition energies.

For reference, in Fig. 6a we show the energies of the lowest
four bands calculated for the same parameters as those used in Fig. 4a except
$\gamma$ = $20^{\circ}$.  Examining Fig. 6a we see that the energy difference
betweeen the third and fourth bands are nearly constant in the region of 
$16 \leq I \leq 24$, while the difference between the lowest and
second-lowest bands is monotonically increasing.
In Fig. 6b calculated $R_i$ values for the bands $b1$ and $b2$ are plotted, 
of which relative values 
are also monotonically changing as a function of $I$.  
Calculated $R_i$-values for the bands $b3$ and $b4$ are shown in Fig. 6c, 
which are drastically and independently changing in contrast to 
the nearly constant energy-difference
between the bands $b3$ and $b4$ shown in Fig. 6a. 
From Figs. 6a, 6b and 6c it is concluded that any two of the four bands, 
$b1$, $b2$, $b3$ and $b4$, hardly form chiral pair-bands.

\section{CONCLUSIONS AND DISCUSSIONS}
Defining chiral pair-bands not only by the degeneracy of the two levels 
with a given $I$ but also by almost equal 
(within $\pm 20$ percent) expectation values of squared components of
three angular-momenta that define chirality, we have explored the presence and
properties of multi chiral pair-bands in odd-odd nuclei, using a particle-rotor
model.  The model is relatively simple, but we believe that it contains basic
elements in physics so that if our model does not at all produce chiral
pair-bands, then there will be presumably little hope 
to obtain chiral pair-bands in
more elaborate models.  

With the parameters of the model that are most favorable for producing chiral
bands it is amazing to see that two lowest $\Delta I = 1$ bands form an almost
perfect chiral pair-bands in the range of $I$ varying the value 
by so much as 10 units.  And, it is also 
possible to obtain the second chiral pair-bands in the
range of $I$ varying the value at least by several units.  
In the region of $I$ the energy difference between the two
$I$-levels of respective chiral pair-bands is one or two orders of magnitude
smaller than a few hundreds keV,  
and mutually
corresponding intraband $M1/E2$ transitions are nearly equal. 
On the other hand, interband $M1/E2$ transitions 
from the second lowest chiral pair-bands to
the lowest ones are found to be too weak to compete with the
intraband $M1/E2$ transitions within the second chiral pair-bands.

Chiral pair-bands, of which the energy difference of the two levels 
with a given $I$ is nearly constant and is the order of a few hundreds keV, have
never been obtained in the present model.  
It is noted that the energy difference comes essentially from the matrix-element
$<IL|H|IR> \neq 0$, which will rapidly decrease as $I$ increases.  
If a pair of bands in odd-odd nuclei, of which the energy
difference is a constant of a few hundreds keV as experimentally reported in the
region of $A \sim 130$, are pinned down to be approximately chiral pair-bands, 
an important and interesting question is what is the origin 
of the energy difference. 
On the other hand, as shown in an example given at the end of section III, in
the present model we sometimes obtain two bands, of which the energy difference
is nearly a constant of the order of a few hundreds keV in a certain range of
$I$.  However, in such cases the examination of the components of three vectors in
(\ref{eq:rsq}) indicates that those two bands are far from being a pair of
chiral bands.  Such two bands may happen to appear  
from the presence
of many close-lying levels (or bands) in odd-odd nuclei.  

Choosing the most appropriate Fermi levels for obtaining chiral pair-bands, 
namely that one quasiproton (quasineutron) expresses almost one hole in the 
high-$j_p$ (high-$j_n$) shell while one quasineutron (quasiproton) represents
one particle in the high-$j_n$ (high-$j_p$) shell, we have looked for the
possibility of obtaining a chiral pair-bands by varying $\gamma$ values. 
It is found that outside the region of $25^{\circ} < \gamma < 35^{\circ}$  
chiral pair-bands following our definition have hardly been obtained.

On the other hand, we have tried numerical calculations by keeping 
$\gamma = 30^{\circ}$ while relaxing the condition that 
a set of one quasiproton and one
quasineutron almost expresses a set of 
one hole and one particle in respective high-$j$
shells.  
Then, the range of $I$ for the occurrence of chiral pair-bands
becomes in general narrower even when the range can ever be obtained.  
For example, simulating somewhat the case of nuclei $^{104,106}Rh$, 
we place the proton Fermi level 
on the second-highest single-particle level of the
$g_{9/2}$-shell and the neutron Fermi level on the second-lowest single-particle
level of the $h_{11/2}$-shell.  Then, it is found that relative values of
energies and $R_i$ of four bands, $b1$, $b2$, $b3$ and $b4$, are monotonically
changing as a function of $I$.  When we take our definition of chiral 
pair-bands we may
barely state that the bands, $b1$ and $b2$, form chiral pair-bands in the
region of $14 \leq I \leq 17$, while the bands, $b3$ and $b4$, form a pair of 
chiral bands for $16 \leq I \leq 18$, though the quality of being chiral
pair-bands is much poorer than that shown in Figs. 4a, 4b and 4c.  
For larger values of $I$ the difference
of zigzag 
pattern (namely odd- and even-$I$ dependence) of $R_i$ values between bands 
$b1$ and $b2$ (and between bands $b3$ and $b4$) becomes so prominent that the
interpretation of the two bands 
as a chiral pair does not work, though their energies
are not so far away from each other.  

An important question is: if  ''the pair'' 
of $\Delta I = 1$ bands observed in odd-odd nuclei in the region of 
$A \approx 130$ and 110 region are not understood 
in terms of chiral pair-bands, we must find what makes the systematic occurrence
of such ''pair'' bands.

The author would like to express her sincere thanks to Professors K. Starosta
and J. Timar 
for showing their interesting data before the publication and 
for stimulating and useful discussions.

\newpage

\noindent
{\bf\large Figure captions}\\
\begin{description}
\item[{\rm Figure 1 :}]
Schematic sketch of the  
selection rule expected for both interbands and intrabands 
$M1(I \rightarrow I-1)$ and  $E2(I \rightarrow I-1)$ transitions between the
two pairs of idealistic chiral bands, ($f1$, $u1$) and ($f2$, $u2$), using
the model discussed in \cite{KSH04}.
The band $u1$ ($u2$) is slightly shifted upward  
from the band $f1$ ($f2$). 
The quantum number $A$ of each level is denoted by $\pm$ sign. 
A band is arranged so that $E2(I \rightarrow I-2)$ transitions are always
allowed within a given band.
The transitions within respective pairs ($f_i, u_i$) are 
expressed by solid-line arrows, while 
those from the second pair-bands to the lowest pair-bands are denoted by
dotted-line arrows.  
There are two possible relations of the quantum number $A$
of the first chiral-pair to that of the second chiral-pair, which are shown in
Figs. 1a and 1b.
See the text for details.
\end{description}

\begin{description}
\item[{\rm Figure 2 :}]
(a) Calculated energies of the lowest four bands for $\gamma$ = 30$^{\circ}$ 
and for the proton (neutron) 
Fermi level placed 
on the lowest (highest) single-particle energy of the $h_{11/2}$-
shell.
(b) Values of $R_1$ and $R_2$ calculated for the lowest two bands.
(c) Values of $j_{pi}$ and $j_{ni}$ calculated for the lowest two bands.  
(d) Values of $R_1$ and $R_2$ calculated for the second-lowest two bands. 
\end{description}

\begin{description}
\item[{\rm Figure 3 :}] 
Parameters are the same as those used in Fig. 2.  The B(M1) values are expressed
in units of $(e \hbar / 2mc)^2$.
(a) Examples of calculated $B(M1;I \rightarrow I-1)$ values of the transitions 
within the first and second pair-bands in the angular-momentum region, 
where the character
of respective chiral pair-bands is confirmed.  
(b) Examples of calculated $B(M1;I \rightarrow I-1)$ values of the transitions
from the second chiral pair-bands to the first ones.
\end{description}

\begin{description}
\item[{\rm Figure 4 :}]
(a) Calculated energies of the lowest four bands for $\gamma$ = 30$^{\circ}$ 
and for the proton Fermi level 
placed on the highest single-particle energy of the $g_{9/2}$-shell while
the neutron Fermi level on the lowest single-particle energy of the
$h_{11/2}$-shell.  
(b) Values of $R_1$ and $R_2$ calculated for the lowest two bands. 
(c) Values of $R_1$ and $R_2$ calculated for the third and fourth bands.
\end{description}

\begin{description}
\item[{\rm Figure 5 :}]
Parameters are the same as those used in Fig. 4.  The B(M1) values are expressed
in units of $(e \hbar / 2mc)^2$.
(a) Examples of $B(M1;I \rightarrow I-1)$ values of the transitions
within the first and second pair-bands, respectively, 
which are calculated in the angular-momentum
region, where the character of respective chiral pair-bands is confirmed. 
(b) Examples of calculated $B(M1;I \rightarrow I-1)$ values of the transitions
from the second chiral pair-bands to the first ones.
\end{description}

\begin{description}
\item[{\rm Figure 6 :}]
Parameters are the same as those used in Fig. 4 except $\gamma$ = 20$^{\circ}$.  
(a) Calculated energies of the lowest four bands.  
(b) Values of $R_1$, $R_2$ and $R_3$ calculated for the lowest two bands. 
(c) Values of $R_1$, $R_2$ and $R_3$ calculated for the third and fourth lowest 
bands.
\end{description}

\end{document}